\documentclass[useAMS,usenatbib]{mn2e}

\ifx\pdfoutput\undefined
\usepackage{graphicx}
\else
\usepackage[pdftex]{graphicx}
\usepackage[pdftex,colorlinks,citecolor=blue,linkcolor=blue,pdfauthor= Matteo~Maturi,pdftitle =Filament~detection,pdfsubject=Cosmology]{hyperref}
\usepackage{epstopdf}
\fi

\usepackage{txfonts}
\usepackage{natbib}
\usepackage{hyperref}
\usepackage{rotating}
\usepackage{enumerate}
\usepackage{url}
\usepackage{color}

\renewcommand{\d}{\mathrm{d}}
\renewcommand{\vec}{\bmath}

\def\aj{AJ}
\def\apj{ApJ}
\def\apjl{ApJ}

\def\apjs{ApJS}

\def\aap{A\&A}

\def\aaps{A\&AS}
\def\mnras{MNRAS}
\def\procspie{Proc. SPIE}
\def\pasp{PASP}

\begin{document}

\title{De-noising the galaxies in the Hubble XDF with EMPCA}

\author[Matteo Maturi]
       {Matteo Maturi\thanks{E-mail: maturi@uni-heidelberg.de}\\
         Universit\"at Heidelberg, Zentrum f\"ur Astronomie, Institut
         f\"ur Theoretische Astrophysik, Philosophenweg 12, 69120 Heidelberg,
         Germany}

\date{\emph{MNRAS, submitted}}

\maketitle

\label{firstpage}

\begin{abstract}

  We present a method to model optical images of galaxies using
  Expectation Maximization Principal Components Analysis (EMPCA). The
  method relies on the data alone and does not assume any
  pre-established model or fitting formula. It preserves the
  statistical properties of the sample, minimizing possible biases.

  The precision of the reconstructions appears to be suited for
  photometric, morphological and weak lensing analysis, as well as the
  realization of mock astronomical images. Here, we put some emphasis
  on the latter because weak gravitational lensing is entering a new
  phase in which systematics are becoming the major source of
  uncertainty. Accurate simulations are necessary to perform a
  reliable calibration of the ellipticity measurements on which the
  final bias depends.

  As a test case, we process $7038$ galaxies observed with the ACS/WFC
  stacked images of the Hubble eXtreme Deep Field (XDF) and measure
  the accuracy of the reconstructions in terms of their moments of
  brightness, which turn out to be comparable to what can be achieved with
  well-established weak-lensing algorithms.
\end{abstract}

\begin{keywords}
  Methods: data analysis, statistical -- Techniques: image processing
  -- Galaxies: general -- Gravitational lensing: weak
\end{keywords}

\section{Introduction}

  Optical imaging is without any doubt one of the main tools to
  investigate galaxies and dark matter through weak and strong
  gravitational lensing. Because of the large available data sets, it
  is crucial to extract all information available in noisy data and to
  simulate images precisely to calibrate the various methods
  and properly deal with possible biases. There is thus a pressing
  need to extract clean galaxy images from data.

  In particular, several studies have shown how all methods used to
  measure the ellipticity of galaxies require realistic simulations
  for their calibration
  \citep{viola11,bartelmann12,refrejer12,2012arXiv1204.5147M,massey13,gorvich16,bruderer16}.
  This issue is becoming pressing because of the stringent
  requirements posed by upcoming wide-field surveys such as the ESA
  Euclid space mission \citep{laureijs09} and the Large Synoptic
  Survey Telescope \citep{kaiser02}, among others. Galaxy models
  based on simple analytical recipes, for example based on the Sersic
  profile \cite{1968adga.book.....S}, have been widely used at this
  end \citep{heymans06,great08,kitching12}. These models have proven
  to well suffice for ground-based observations, but more accurate
  simulations are now necessary to include complex morphologies to
  account for spiral, irregular and cuspy galaxies.  Also in the
  strong gravitational lensing regime, accurate galaxy models are now
  needed to investigate the use of substructures of strongly magnified
  galaxies to better constrain the mass distribution of lenses such as
  for example galaxy clusters \citep{meneghetti08,zitrin15}.

  For these reasons, galaxies observed with HST have been modelled
  with shapelets \citep{refregier03,refregier03b,massey03,massey07}
  to achieve noise-free images. Even if this approach deals with complex
  morphologies, artifacts may arise because of the oscillatory
  behavior of shapelets. Moreover, also smooth galaxies such as for example
  ellipticals are not very well reconstructed by this approach because
  of their slope, which is not well compatible with a Gaussian,
  which the Hermite polynomials in the shapelets are derived from. Also, image
  cut-outs have been extracted from HST data
  \citep{rowe14,mandelbaum15}, but these stamps are affected by the
  instrumental noise limiting their applicability.

  In this paper, we present a method to retrieve and reconstruct clean galaxy
  images in a model-independent way, which also preserves the
  statistical properties of the reconstructed sample. We do this using
  Expectation Maximization Principal Components Analysis
  \cite{bailey12}, which we use to derive a set of orthonormal basis functions
  optimized for the specific data set to be processed. Other
  studies used standard PCA \citep{Jolliffe86} to model elliptical
  galaxies \citep{joseph14J}. However, these are galaxies with smooth
  morphology, and this method cannot deal with weights and missing
  data. In contrast, the procedure discussed in this work allows us to
  process astronomical images with masked areas and pixel-dependent
  variance, and it allows us to introduce regularization terms to be used
  when deriving the principal components and impose smoothness on the
  basis.

\begin{figure}
  \centering
  \includegraphics[width=1.0\hsize]{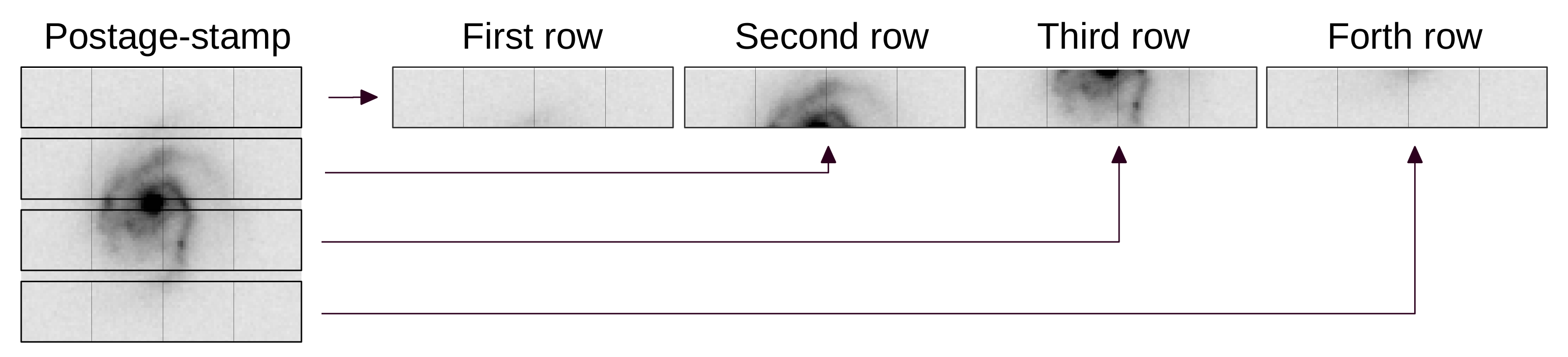}
  \caption{This schematic representation shows how a postage-stamp of
    a galaxy image is rearranged in form of a vector, $\vec{d}_i$. All
    rows of pixels in the matrix composing the image are simply
    concatenated.}
  \label {fig:subaru-train-cov}
\end{figure}

\begin{figure*}
  \centering
  \includegraphics[width=0.33\hsize]{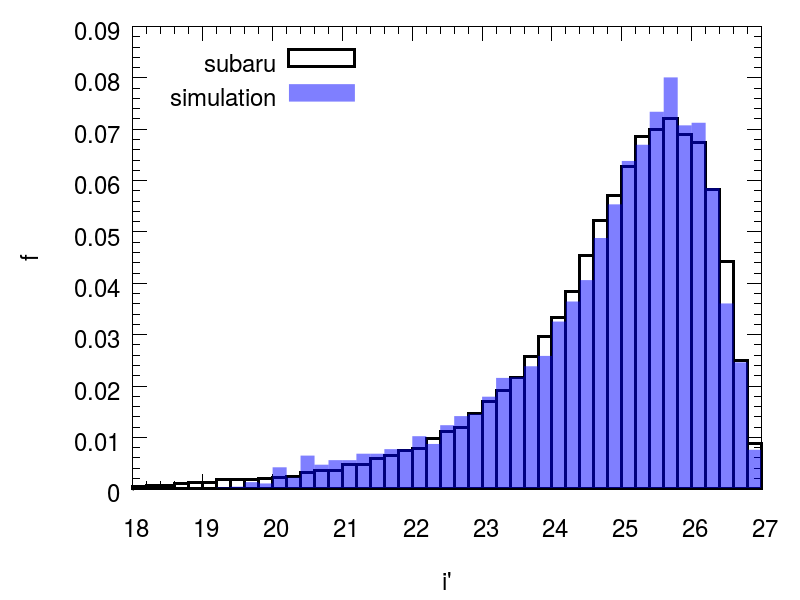}
  \hfill
  \includegraphics[width=0.33\hsize]{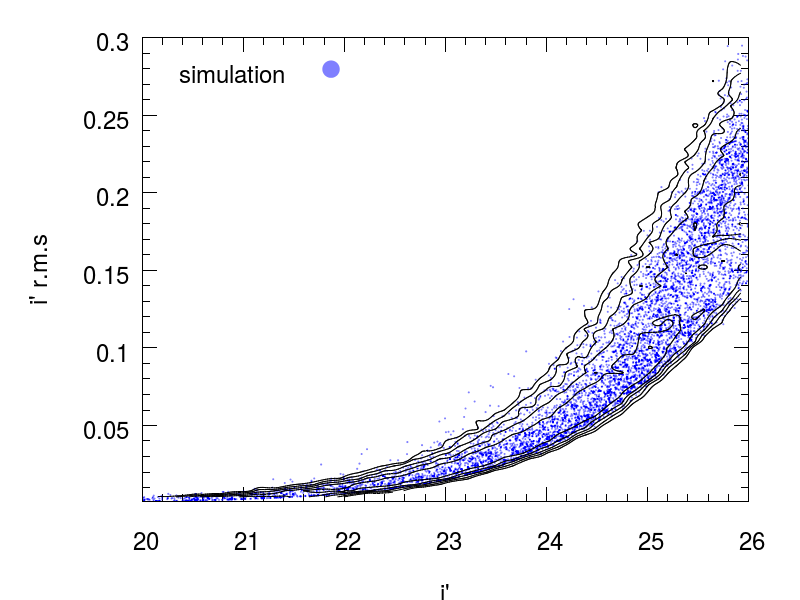}
  \hfill
  \includegraphics[width=0.33\hsize]{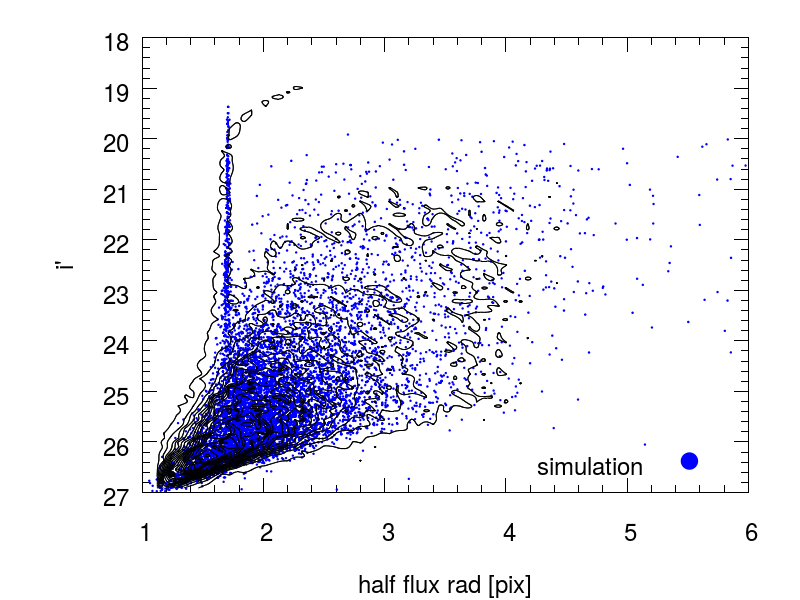}
  \caption {Statistical properties of the sources detected in the
    simulation and reproducing a Subaru image with 20 minutes of
    exposure time. A comparison with the real data is provided. From
    left to right, the panels show the magnitude distribution of the
    objects, their photometric error as a function of magnitude, and
    their size versus magnitude relation.
  }
  \label{fig:sim-stat}
\end{figure*}

As a test case, we analyzed $7038$ galaxies extracted from the
\cite{rafelski15} catalog with redshift up to $z<4.0$ and maximum
magnitude up to $m_{F775W}<30$. The catalog contains all photometric
information including the photometric redshifts of the objects. We
modeled these galaxies in all 5 optical bands extracted from the
Hubble eXtreme Deep Field \citep[XDF hereafter,][]{illingworth13}. We
tested the quality of the models by comparing their moment of
brightness against those measured with the weak-lensing Shapelens
library \citep{2011MNRAS.412.1552M}. Moreover we showed how to use
these models to construct realistic simulations of astronomical
images.

The structure of this paper is a follows: in Sec.~(\ref{sec:empca}), we
derive the EMPCA which are then used in Sec.~(\ref{sec:models}) to
create the models of the galaxy images. The description of the
analysis of the XDF data set in Sec.~(\ref{sec:models}), a simple sky
simulation based on our models is presented in
Sec.~(\ref{sec:simulations}), and the conclusions are given in
Sec.~(\ref{sec:conclusions}).

\section{A linear model for galaxy images}\label{sec:empca}

In this section, we discuss how to model the images of individual
galaxies to obtain a noise-free reconstruction. Let us now consider
the case in which we have one single object placed in the center of a
postage-stamp. This cut-out can be modeled as
\begin{equation} \label{eqn:datamodel}
  d(\vec{x})=g(\vec{x})+n(\vec{x})  \;,
\end{equation}
where $g(\vec{x})$ is the object contribution we are interested in,
$n(\vec{x})$ the one of the noise (e.g. photon noise, read-out noise,
dark current), and $\vec{x}\in\mathbb{R}^{2}$ denotes the position in
the image. For simplicity, we assume the noise to be uncorrelated with
standard deviation $\sigma$, defined by $\langle n_i\,n_j \rangle
=\sigma^2\,\delta(i-j)$. This 2-dimensional image $d(\vec{x})$,
consisting of $n=n_x \times n_y$ pixels, can be represented as a data
vector $\vec{d} \in \mathbb{R}^{n}$ whose elements $d(x_i)=d_i$ are
the intensities of the $i_{th}$ pixels. A visual impression of how
pixels are rearranged in a vector is shown in
Figure~(\ref{fig:subaru-train-cov}).

The most general linear model to describe this data element is
\begin{equation} \label{eqn:model}
  \tilde{d}(\vec{x})=\sum_{k=1}^M a_k \phi_k(\vec{x}) \;,
\end{equation}
where $\vec{\phi}_k\;$ is a collection of $M$ vectors,
$\left\{\vec{\phi}_k \;\in\; \mathbb{R}^{n} \;\mid\; k=1,...,M\right\}$. The
goal now is to define an optimal set of vectors $\phi_k$ capturing the
relevant signal and sort them depending on their information content
(power) such that each vector contains more information than the
following one.

Once this is achieved, the sm can be split into two terms
\begin{equation}\label{eq:model-split}
  d(\vec{x})=\sum_{k=1}^M a_k \phi_k(\vec{x}) + \sum_{M+1}^{n} a_k \phi_k(\vec{x}) = \tilde{g}(\vec{x}) + \tilde{n}(\vec{x}) \;,
\end{equation}
where now $\tilde{g}(\vec{x})$ is the model of the object we are
interested in, and $\tilde{n}(\vec{x})$ a term containing most of the
noise and a small, and hopefully negligible, amount of information.
The number of components, $M$, fixes the amount of information which
is going to be kept in the model and the amount of noise which is
going to be suppressed. Some information loss is inevitable, otherwise
one could fully recover the real image of the object which is
obviously an impossible task. A common way to achieve this
decomposition is provided by the Principal Component Analysis
\cite[PCA hereafter,][]{Jolliffe86} which takes advantage of the
entire data set, i.e. the postage-stamps of all galaxies in the
sample, $\left\{\vec{d}_j \;\in\; \mathbb{R}^n \;\mid\;
j=1,...,s\right\}$, and consists in finding the set of vectors,
$\vec{\phi}_k\; \in\;\mathbb{R}^{n}$, minimizing the quantity
\begin{equation}\label{eq:chisq}
  \chi^2 = \sum_{ij}^{n,s} \left(d_{ij}-\sum^n_{k=1}a_{kj}\phi_{ki}\right)^2 \;.
\end{equation}
In other words, we are looking for the model based on all coefficients,
$a_{jk}$, and vectors, $\vec{\phi}_k$, which best fit all images at
once. Here and throughout the paper, the index $i$ runs over the
number of pixels, $j$ over the number of galaxies, and $k$ over the
number of components which, in the case of the PCA, is equal to the
number of pixels. The coefficients of the $j_{th}$ galaxy are derived
with the scalar product $a_{kj}=\sum_id_{ji}\phi_{ki}$ or by linear
fitting if needed. Equation~(\ref{eq:chisq}) can be easily generalized
to account for correlated noise.

\begin{figure}
  \centering
  \includegraphics[width=0.98\hsize]{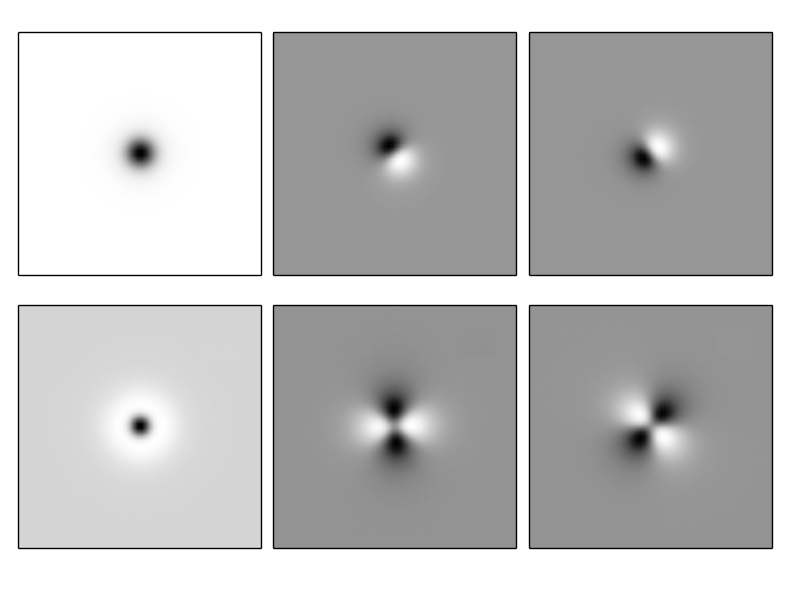}
  \caption{First six principal components, $\vec{\phi}_i$, derived
    from the noisy simulated image. The components are rearranged as
    images. From the data emerge the main features related to dipolar
    and quadrupolar structures as well as radially symmetric ones.}
  \label{fig:subaru-basis}
\end{figure}

\begin{figure}
  \centering
  \includegraphics[width=0.95\hsize]{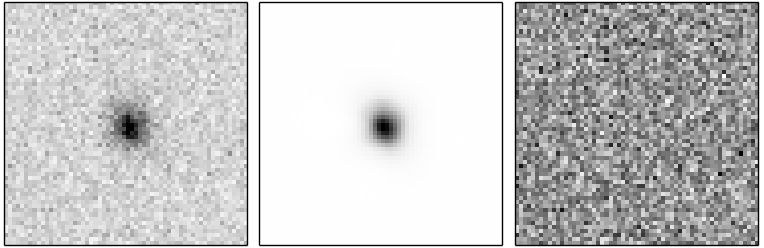}
  \caption{One of the galaxies randomly extracted from the simulation
    for which we show the signal and noise splitting: in the left
    panel the original postage-stamps, $\vec{d}_j$, the model,
    $\vec{\tilde{g}_j}$ in the center and the residuals,
    $\vec{\tilde{n}_j}$ on the right.
}
  \label{fig:subaru-object}
\end{figure}

Usually, the principal components are found by diagonalizing the
centered covariance matrix of the data and ordering its eigenvectors
by decreasing eigenvalues, $\lambda_k$. Another way to find
the solution for the minimum of Equation~(\ref{eq:chisq}) is through
the Expectation Maximization Principal Components Analysis (EMPCA
hereafter). This is an iterative algorithm which consists in finding
one basis component at a time. To find the first component $\vec{\phi}_1$, we
start with a random vector (all its $n$ components have a random
value) with which we compute the first coefficient of all objects
through the scalar product previously defined,
$a_{1j}=\sum_id_{ji}\phi_{1i}$. These coefficients are then used to
update $\vec{\phi}_1$,
\begin{equation}
  \phi_1^{new}(\vec{x}) = \frac{\sum_j^s a_{1j} d_j(\vec{x})}{\sum_j^s
    a_{1j}^2} \;,
\end{equation}
which is renormalized for convenience and which now will fit the data
set better. The refined $\vec{\phi}_1^{new}$ vector is subsequently
used to compute a new set of $a_{1j}$ coefficients necessary for the
next iteration, and so on. This procedure is converging toward the
only absolute minimum of the $\chi^2$ function, as demonstrated by
\cite{srebro03}. In our case, this iterative process is stopped when
the variation in the principal component, $\vec{\phi}_1$, is smaller
that a certain value, $|\Delta\phi_1|<\epsilon$, which we set to
$\epsilon = 10^{-6}$. The next principal component,
$\vec{\phi}_{k+1}$, is found by applying the same procedure to that
part of the signal which has not been captured by the previous $k$
components,
\begin{equation}
  \phi_{k+1}^{new}(\vec{x}) = \frac{\sum_j^s a_{k+1 j}\, \tilde{n}_{j}(\vec{x})}{\sum_j^s a_{k+1 j}^2} \;.
\end{equation}
Here, $\tilde{n}_{j}(\vec{x}) = d_j(\vec{x})-\tilde{g}_{j}(\vec{x})$
is the residual part of the signal for which this component is
evaluated, and $\tilde{g}_{j}(\vec{x})=\sum_{l=1}^k a_{lj}
\phi_l(\vec{x})$ is the updated model of the $j-th$ galaxy. In analogy
to the derivation of the first component, the coefficients of the
$j-th$ galaxy, $a_{k+1\,j}=\sum_id_{ji}\phi_{k+1\,i}$, are based on
the previous iteration, and $\vec{\phi}_{k+1}^{new}$ is renormalized
each time. With this procedure we compute at the same time the
principal components, $\vec{\phi}_k$, the noise, $\tilde{\vec{n}}_j$,
and the signal estimate, $\tilde{\vec{g}}_j$, we are aiming at (see
Equation~\ref{eq:model-split}). The procedure ensures the
orthogonality of the principal components. The further advantages of
this method with respect to a geometrical interpretation of the
principal components are that it allows to take advantage of weights,
masks and the noise covariance by including them in the $\chi^2$ function,
and to impose further conditions on the basis such as for example a
regularization term to obtain smooth basis components \citep{bailey12}.

\begin{figure}
  \centering
  \includegraphics[width=0.98\hsize]{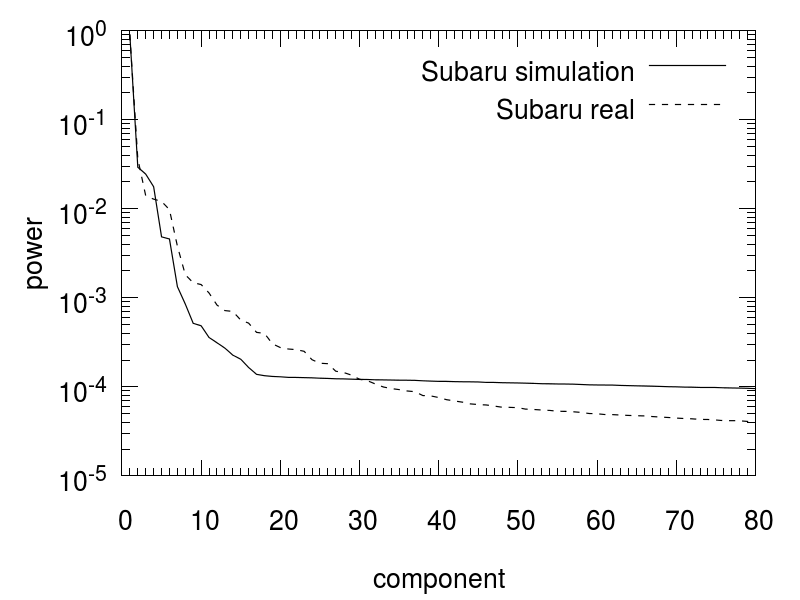}
  \caption{Power of the principal components resulting for the
    simulation with Sersic galaxies (continuous line) and for the real
    Subaru image mimicked by the simulation (dashed line). The power
    has been normalized with respect to the total variance,
    $\sigma^2=\sum_k^n\lambda_k$. In both cases, their amplitude drops
    rapidly with the order. }
  \label{fig:subaru-eigenvalues}
\end{figure}

Here, we have derived smooth principal components by
bilateral smoothing during their construction. This is an edge-preserving
algorithm with an adaptive smoothing scale which allows to
leave those regions of the basis unaffected which show steep gradients that
would be blurred otherwise. The bilateral
smoothing implemented is based on the product of two Gaussians, one acting on
angular scales as it would do a normal convolution and another based on
the local luminosity gradient such that areas with ``sharp'' features
do not get smoothed (if the gradient is small the angular convolution
takes the lead, while it is made ineffective otherwise).

Alternatively, we used
a different regularization scheme base on a Savitzky–Golay filter
and obtained similar results. It is important to introduce this
regularization during the basis construction and not thereafter (by
smoothing the final basis or the models) to preserve the
orthonormality of the basis, the information present in the data, and
to make the basis more stable against noise
fluctuations.

Additionally, we reduced the scatter in the galaxy
morphologies by rotating the input images by multiples of $\pi/2$ such
as to have their position angle range between 0 and $90$ degrees. We
chose such discrete rotations instead of aligning the galaxies
along their major axes to avoid the introduction of any pixel
correlations.

\begin{figure*}
  \centering
  \includegraphics[width=0.48\hsize]{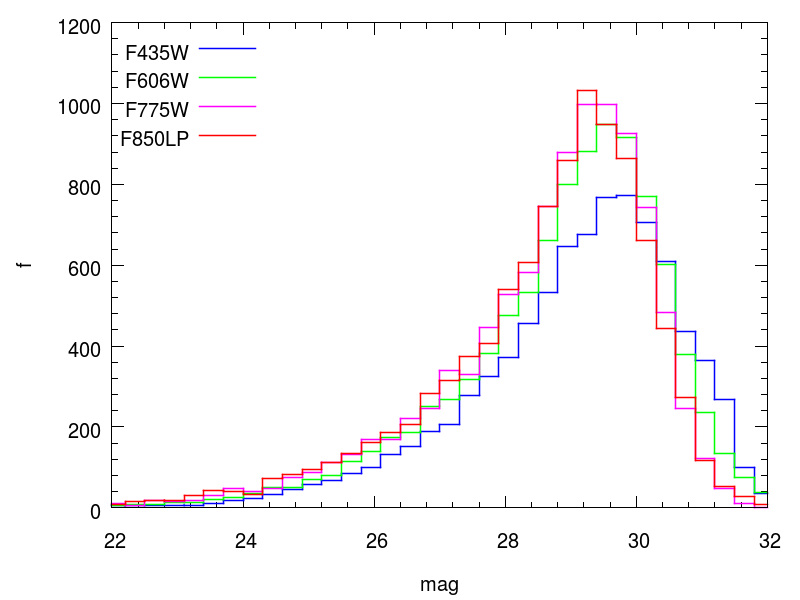}
  \includegraphics[width=0.48\hsize]{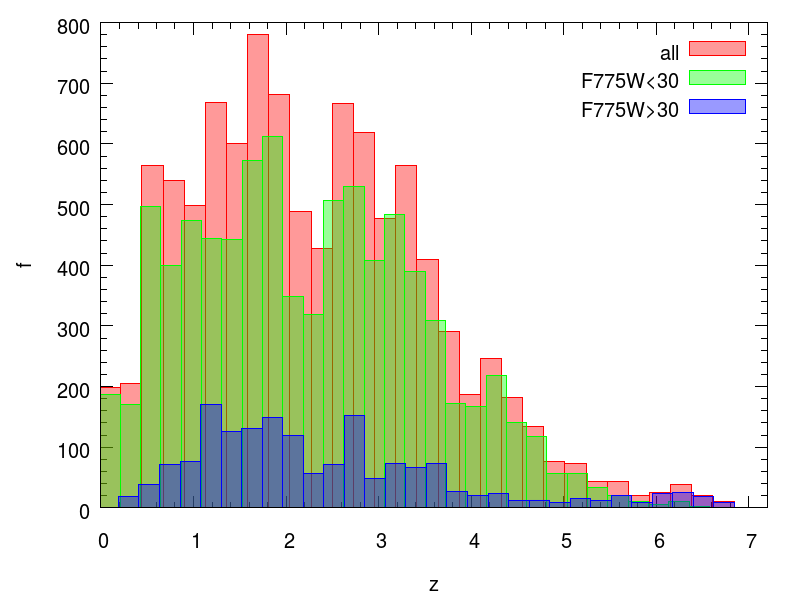}
  \caption{Magnitude (left panel) and redshift
    distributions (right panel) of the sources with a maximum
    $\chi^2_{mod}$ of 4 for the BPZ redshift. The right panel shows
    the number of galaxies in the bins split according to their F775W
    magnitude.}
  \label {fig:cat-photo-z}
\end{figure*}

In the following, we focus on the caveats of this approach which one
should keep in mind to properly use it. First of all, the basis is
derived from noisy data, and this will of course impact on
them. This is relevant at higher orders where the EMPCA have to deal
with smaller and less prominent features approaching the noise regime,
as is clear from the discussion of Eq.~(\ref{eq:model-split}). Second,
the number of galaxies to produce the basis is limited, and a
small sample will not be representative of the entire
population. On the other hand, wide field surveys provide us with large data
sets which even allow to split the sample into subsets of galaxies with
similar properties (such as for example size or ellipticity) and
further reduce the data scatter for a better optimization of the
basis.

It is also preferable to compute the basis for the sample which has to
be processed and not on another independent set of images even if with
compatible size and quality. After all, the basis components are evaluated by
finding the optimal model of the specific data at hand. Third, the
regularization used to impose smoothness on the basis
might decrease the level of high frequency features
present in the data, if one wants to
use it at all. In any case, one can always decide and adapt the
regularization which better suits the case at hands.

One final remark about the point spread function (PSF): The PSF is not
an issue for this approach because it aims at the image as it
is. However, some care has to be taken if the models are going to be
used for weak-lensing purposes because the PSF ellipticity and
distortions may leave an imprint on the derived basis and thus on the
models based on them. It is easy to cope with this issue with the
procedure adopted in this work. In fact, the rotation applied to the
postage-stamps prevents any isotropy which my be induced by the PSF.

\section {Defining the number of components}\label{sec:sn-separation}

As discussed in Section~(\ref{sec:empca}), the number of basis components,
$M$, defines the amount of noise which is going to be suppressed and
the amount of information which is going to be kept in the model. Its
value depends on the specific task we have at hands. Various schemes
for the definition of $M$ have been proposed in the literature, for
instance through the analysis of the scree graph
\citep{cattell66,cattell77} or the log-eigenvalue diagram (LEV)
\citep{Farmer71,maryon79,beltrando90}. These approaches are not very
stable because they largely depend on specific features of these
diagrams which may not be well defined in certain cases. Other
approaches rely on a $\chi^2$ approach,
\begin{equation}\label{eq:chisq_new}
  \chi_j^2 = \sum_{i=1}^n (d_{ij}-\tilde{g}_{ij})^2 \,
\end{equation}
applied to each individual galaxy (not the overall set!) and which
better suits our needs \citep{ferre90}. Below, we discuss two criteria
aiming at different goals.
\begin{enumerate}

\item {\it Defining $M$ to minimize the model variance}: if we search
  for the model $\tilde{g}$ with the minimum variance, it is necessary
  to minimize the number of basis components. Here we seek for the
  smallest $M$, which will be specific for each individual object, by
  including one component at a time until a certain convergence
  criteria is reached, for instance until we obtain a reduced $\chi^2$
  close to unity or until the $\chi^2$ is not changing by more than a
  certain threshold. Formally this criterion reads $\left\{\,
  \mbox{min}\{k\} \in \mathbb{N} : \chi^2_k -\chi^2_{k+1}\le t
  \,\right\}$, where $t$ is the threshold to be set. A note of caution
  is in order here: by construction, the $\chi^2$ is monotonically
  decreasing with the order (at the order $n$ the $\chi^2$ will be
  zero) and under- or over-fitting might be an issue.  Moreover, the
  basis components are sorted by their information content (the higher
  the order, the less relevant is the component), but this sorting is
  based on the statistics of all objects in the sample and may not be
  proper for a specific object. It may happen that for a specific
  object, one of the higher-order components is more relevant than
  lower-order components, and the convergence process may stop before
  this component is reached.

\item {\it Defining $M$ to maximize model fidelity}: for ensuring not
  to miss any valuable signal, it is necessary to include all
  components, for instance by visually inspecting the basis or by
  finding the $M$ for which the expectation value of the global
  $\chi^2$ is the minimum. This $M$ can not be evaluated directly but
  it can be approximated by
  \begin{equation}\label{eq:expectationval}
    \hat{f}_q=\sum_{k=q+1}^n \hat\lambda_k + \sigma^2
    \left[ 2nq-n^2+2(n-q)+4\sum_{l=1}^q\sum_{k=q+1}^n\frac{\hat\lambda_l}{\hat{\lambda}_l-\hat{\lambda}_k} \right] \;,
  \end{equation}
  under the assumption of uncorrelated noise. Here, $\lambda_k$ is the
  power related to the $k-th$ component. The variance can be estimated
  as $\sigma^2_q=1/(n-q)\sum_{k=q+1}^n\lambda_k$ and
  $\hat{\lambda}_k=\frac{n-1}{n}\lambda_k$ \citep{ferre90}. With these
  criteria, $M$ is the same for all objects because it is based on the
  statistics of the entire data sample, in contrast to what we
  discussed in point (i) where the $\chi^2$ was evaluated for each
  individual object. This approach returns the highest ``fidelity''
  because the largest number of sensible components is used.
\end{enumerate}

\section {Simulation of ground base observations}\label{sec:simulation_ground}

To test the quality of the model reconstruction, we produced a set of
simulations with {\it EasySky}. {\it EasySky} allows to use any object
contained in a postage-stamp image, Sersic galaxies, galaxies with a
single or double Sersic components (bulge plus disc) in the same
fashion as the Great08 \citep{great08,kitching12} simulations, and stars
with a Moffat profile and arbitrary ellipticity (if stars are
to be included). The objects can be displaced in various ways: randomly across
the whole field-of-view, on a regular grid with stars on one side and
galaxies on the other, or on a regular grid but with the stars located
equidistantly from the other galaxies. The galaxies
can be (1) randomly rotated, (2) kept with their semi-major axes aligned
with one direction, (3) within the same quadrant, or (4) produced in
pairs with angles rotated by 90 degrees with respect to each
other. The latter configuration is a useful feature for weak-lensing
calibrations. The fluxes and sizes of both galaxies and stars can be
kept fixed or randomly distributed following a given luminosity
function to better re-sample the data to be simulated. The image can be
convolved with an arbitrary kernel, and a simple shear distortion can
be applied to the galaxies.

\begin{figure}
  \centering
  \includegraphics[width=1.0\hsize]{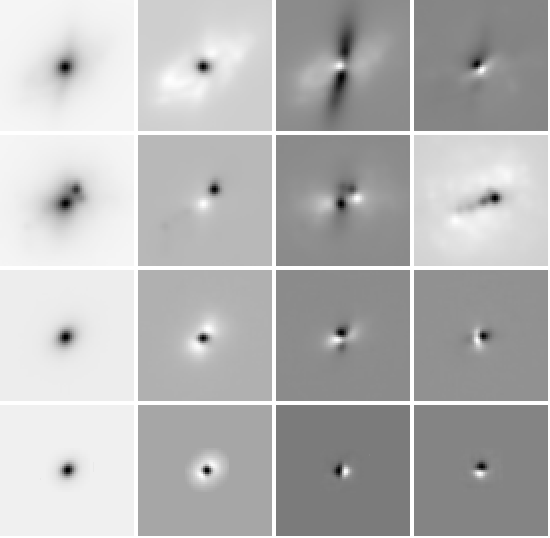}
  \caption {The first four principal components for three different
    sets of galaxies. The upper, middle and bottom rows refer to
    galaxies with semi-major axis, $A$, within the range $5<A<7$,
    $18<A<20$, $50<A<60$, and $70<A<80$ pixels, respectively. The
    larger the galaxies are, the larger is the typical size of the principal
    component and the larger is their morphological complexity.}
  \label{fig:basis}
\end{figure}

Here, we describe a simple but quite realistic synthetic image which
we used to show the signal-to-noise splitting of
Eq.~(\ref{eq:model-split}). We included $10\,000$ galaxies located on
a $100\times100$ regular grid but with the centroid displaced by a
random shift within $1.5$ pixels. The galaxies are characterized by a
Sersic profile \citep{1968adga.book.....S} with a fixed index
$n=2$. They are convolved with a PSF described by a Moffat function
with $\beta=4.8$, $\mbox{FWHM}=4.45$ pixels \citep{moffat69}, and a
complex ellipticity $g_{PSF}=-0.019-i\,0.007$, adapted to the fiducial
values adopted in Great08 \citep{great08}. The noise variance, the
source fluxes, scale radii and ellipticities have been randomly
distributed such as to resemble those of a stacked image obtained with
the OmegaCam mounted on the Subaru telescope and 20 minutes of total
exposure time in the $i^\prime$ filter. A comparison of the magnitude
distribution, the photometric errors and the size-magnitude relation
between the simulated images and a real Subaru image is shown in
Fig.~(\ref{fig:sim-stat}).

The simulation has been processed as it would be with a real
image, i.e. detecting the sources, separating galaxies from stars
with SExtractor \citep{1996A&AS..117..393B}, and creating
postage-stamps sized $60\times60$ pixels for each object based on
the measured astrometry. In this case, we did not apply any rotation
to the galaxy images for simplicity. This will be done in the more
sophisticated simulation discussed in
Sec.~(\ref{sec:simulations}) below. The principal components,
$\vec{\phi}_i$, obtained for this sample are shown in
Fig.~(\ref{fig:subaru-basis}), where for visualization purposes only
we inverted the process sketched in
Fig.~(\ref{fig:subaru-train-cov}) to rearrange the vectors in form
of an image.

It is interesting to note how the data deliver principal components
with radially symmetric profiles ($w_1$ and $w_4$), dipolar
($w_2$ and $w_3$), and quadrupolar ($w_5$ and $w_6$) structures as
well. Higher modes show hexapoles and more complex structures. It is
easy to interpret these shapes: for instance the circularly symmetric
components take care for the average brightness profile, while the dipolar ones
account for a large fraction of the object's ellipticity.

In Fig.~(\ref{fig:subaru-eigenvalues}), we show the power of the
principal components normalized by the total variance,
$\sigma^2=\sum_k^n\lambda_k$. The continuous line represents the
simulation, and the dashed line the real SUBARU image mocked by the
simulation. In both cases, the power of the components drops rapidly
with their order, and the same kinks are visible in the curves.
The drop in power is
less dramatic for real data because of the more complex
morphology of the galaxies. At orders higher than $15$, there is
a clear plateau for the simulated images because here we enter
the regime of uncorrelated random noise containing no more features.
The real image lacks such a plateau because in this case the
noise is correlated because the image results from stacking.

\begin{figure}
  \centering
  \includegraphics[width=1.0\hsize]{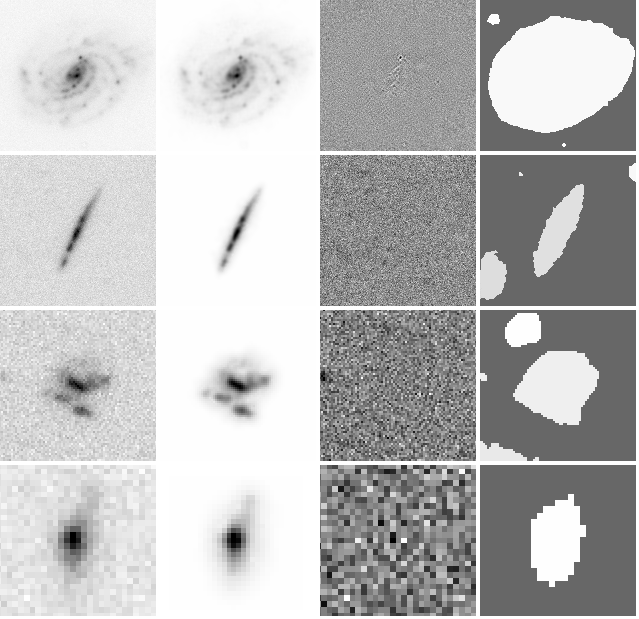}
  \caption {Four galaxies extracted from the XDF sample. From left to right, the columns
    show: the original image (with other
    objects present in the postage-stamp already removed),
    the model generated with the EMPCA, the model residuals, and the
    segmentation of the image we derived from the model. The image of
    the smaller galaxies has been enlarged for visualization
    purposes.}
  \label{fig:models}
\end{figure}

An example for the signal-to-noise splitting, $d(\vec{x})=
\tilde{g}(\vec{x}) + \tilde{n}(\vec{x})$, discussed in
Sec.~(\ref{sec:empca}), is displayed in
Fig.~(\ref{fig:subaru-object}) where, from left to right, we show
the original data $\tilde{d}(\vec{x})$, the model
$\tilde{g}(\vec{x})$, and the residuals $\tilde{n}(\vec{x})$. The
maximum order $M$ has been determined with the criterion $\left\{\,
\mbox{min}\{k\} \in \mathbb{N} : \chi^2_k -\chi^2_{k+1}\le t
\,\right\}$, setting $t=0.005$. As expected, the galaxy model is
compact, i.e. it vanishes at a certain distance form the center of the
galaxy, most of the noise is removed from the image, and the residuals
are fully uncorrelated.

A more quantitative assessment of the reconstruction quality, based on
the brightness moments of the images, is discussed in
Sect.~(\ref{sec:simulations}) where we deal with galaxies with
complex morphologies.

\section{Modeling the XDF galaxies}\label{sec:models}

We now come to the full analysis of real data. Here, we processed the
ACS/WFC stacked images of the Hubble eXtreme Deep Field \citep[XDF,
  hereafter, see][]{illingworth13} which covers an area of $10.8$
arcmin$^2$ down to the $\sim30$ AB magnitude ($5\sigma$). The images
are drizzled with a scale of $0''.03$ pixel$^{-1}$ and have been
obtained with the F435W, F606W, F775W, F814W, and F850LP filters for a
total exposure time of $1177$ks. We used all objects listed in the
UVUDF catalog which are classified as galaxies \citep{rafelski15}. To
avoid artifacts and truncated objects, we discarded those objects in
the areas affected by the ghosts and halos of stars or close to the
edges of the survey. In this way, we selected 8543 galaxies from an
effective area of $9.20$ arcmin$^2$. We further cut the sample by
rejecting all galaxies with an F775W magnitude larger than $30$,
ending up with 7038 objects. The redshift and magnitude distributions
of the sources with a maximum $\chi^2_{mod}$ of 4 for the BPZ redshift
are shown in Fig.~(\ref{fig:cat-photo-z}).

\begin{figure}
  \centering
  \includegraphics[width=1.0\hsize]{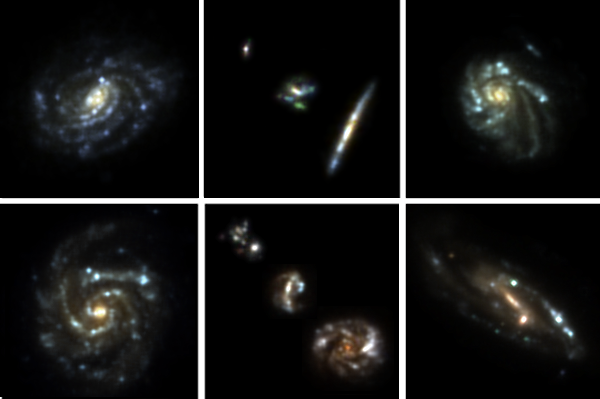}
  \caption {A collection of galaxy models obtained with the EMPCA and
    based on the F435W, F606W, and F775W filters. The first and second
    top panels show the same objects displayed in
    Fig.~(\ref{fig:models}). The size of each box is of $5.4$
    arcsec.}
  \label{fig:color-stamps}
\end{figure}

\begin{figure}
  \centering
  \includegraphics[width=1.0\hsize]{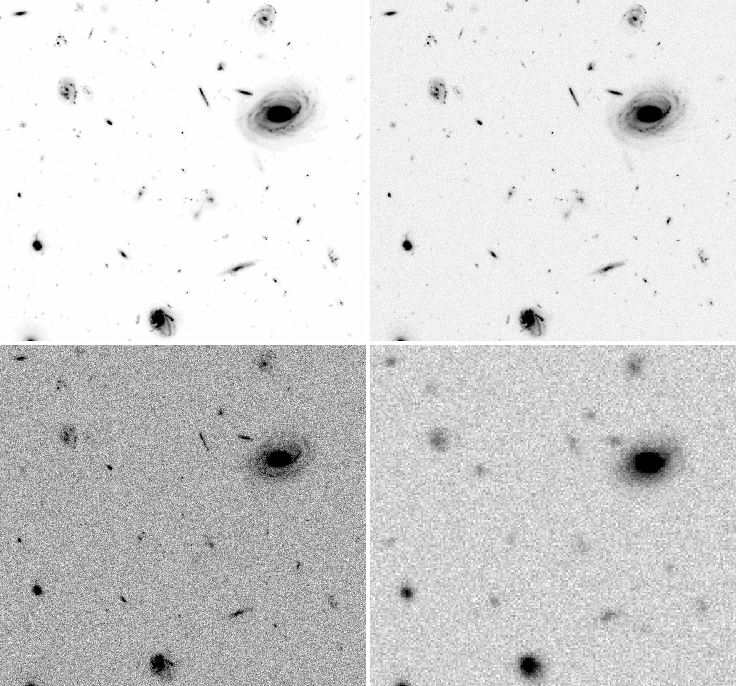}
  \caption {Four simulations realized with {\it EasySky} displaying a
    field of $0.65\times0.65$ arcmin$^2$ observed with the F775W
    filter without noise (upper left panel), with a XDF like noise
    (upper right panel), a noise equivalent to one HST orbit (bottom
    left), and a CFHTLens-quality image for a seeing of $0.7$ arcsec
    (bottom right panel). }
  \label {fig:comparison-clean-dirty}
\end{figure}

When computing the principal components, we split the sample in groups
of galaxies with similar size, and evaluate the basis for each of
these subsamples separately. This is to obtain a collection of basis
sets, one per sub-sample, which is optimized for galaxies with that
specific size. If we used all galaxies at once, the features
captured by the EMPCAs will be distributed on a larger number of
components because then the range of sizes they have
to reproduce will be larger.

To further reduce the amount of scatter in the data, we
rotated the galaxies by $90$ deg, whenever necessary, to align them
within the same quadrant. We did not align the galaxies' major axes
along the same direction to avoid the introduction of additional
correlation among the pixels for a marginal
improvement that could barely be justified.

We finally take advantage of all bands by including all
of them in the training sets. This further reduces the noise and
enriches the number of features which can be reproduce with the same
basis set. In Fig.~(\ref{fig:basis}), we plot the first four
principal components computed for three different sub-sets of galaxies
with semi-major axis, $A$, within the ranges $5<A<7$, $18<A<20$,
$50<A<60$, and $70<A<80$ pixels for the first, second, third and forth
rows, respectively. The semi-major axes, $A=\sqrt{I (1-e) / \pi}$, was
derived from the isophotal area, $I$, and the object
ellipticity, $e$, which turned out to be a suitable choice for our
pourpose. As expected, the larger the galaxies in the training sample are,
the larger is the typical scale of the EMPCA. Additionally, one can see
that larger galaxies show principal components with more complex
features because of their larger variety in morphology and
substructure.

Having computed the basis for each sub-sample, we use them to create
the galaxy models as discussed in Sec.~(\ref{sec:empca}).  Since some
of the variance due to the noise is still present in the model, which
is unavoidable in general, we set all pixels to zero with amplitude
less than $t=\sigma/5$, where $\sigma$ is the pixel variance of the
original image.  This is to avoid such areas of the image which, in
fact, do not show any evidence of signal. The code to perform the
overall analysis is called {\it EasyEMPCA}.

\begin{figure*}
  \centering
  \includegraphics[width=0.48\hsize]{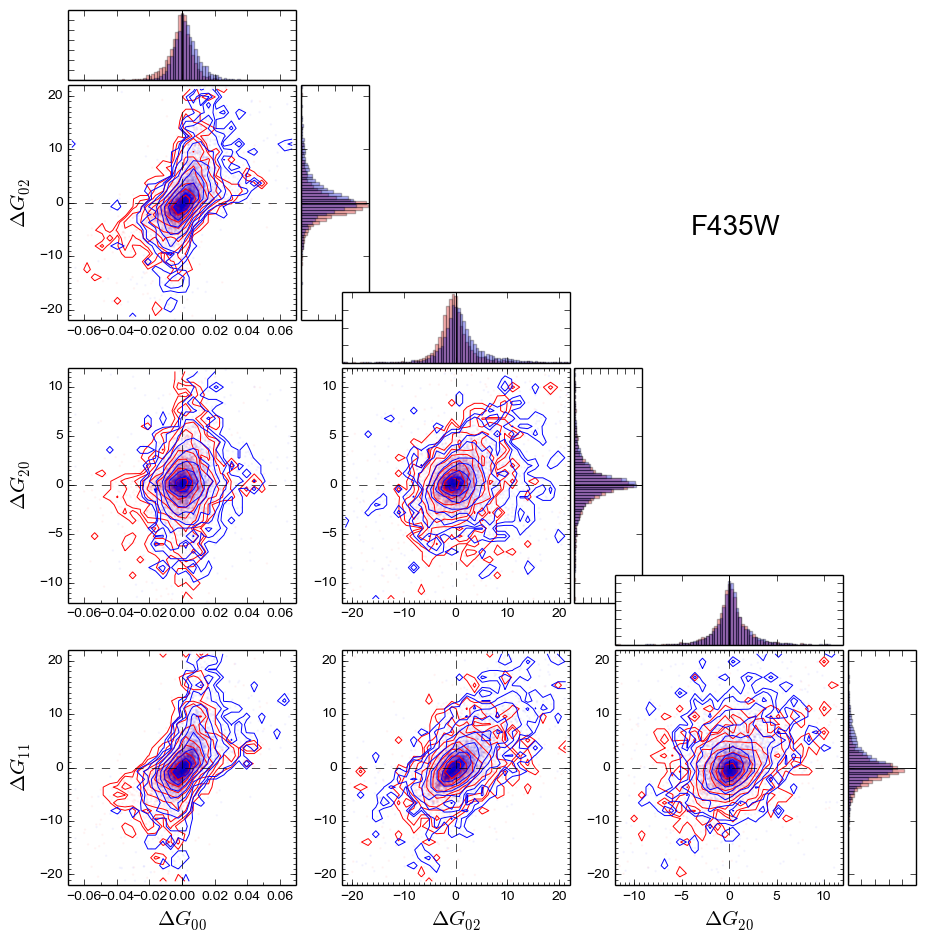}
  \includegraphics[width=0.48\hsize]{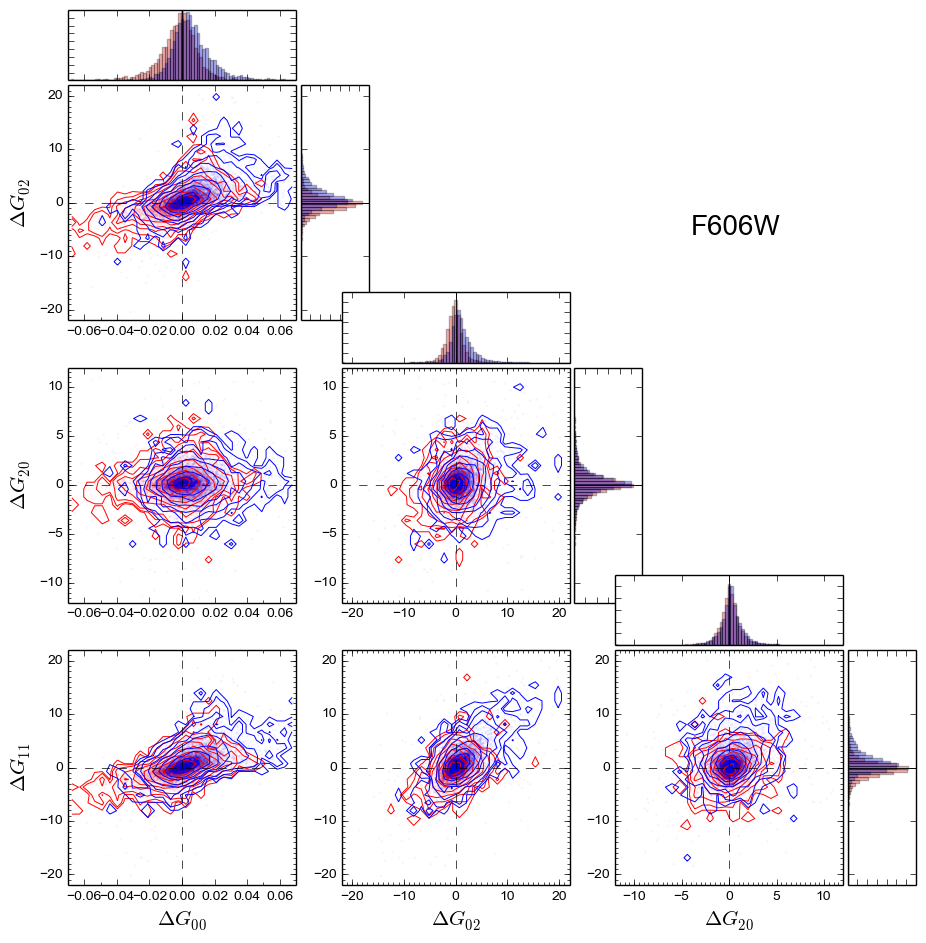}
  \includegraphics[width=0.48\hsize]{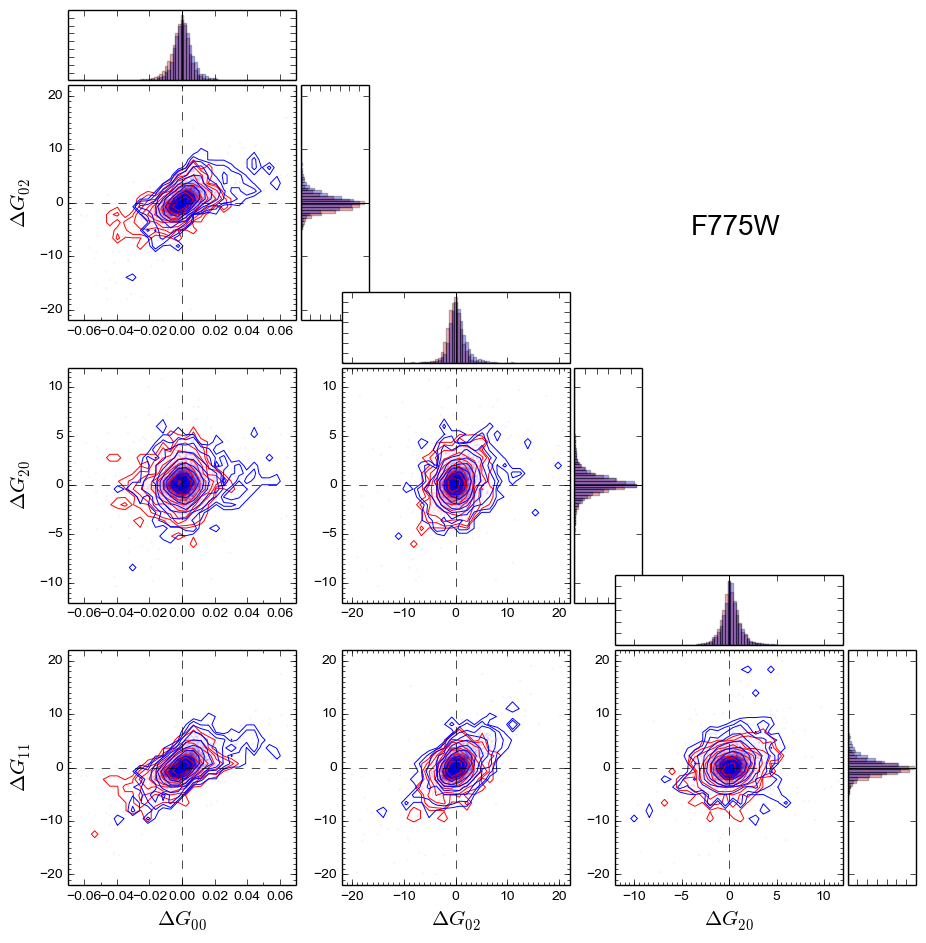}
  \includegraphics[width=0.48\hsize]{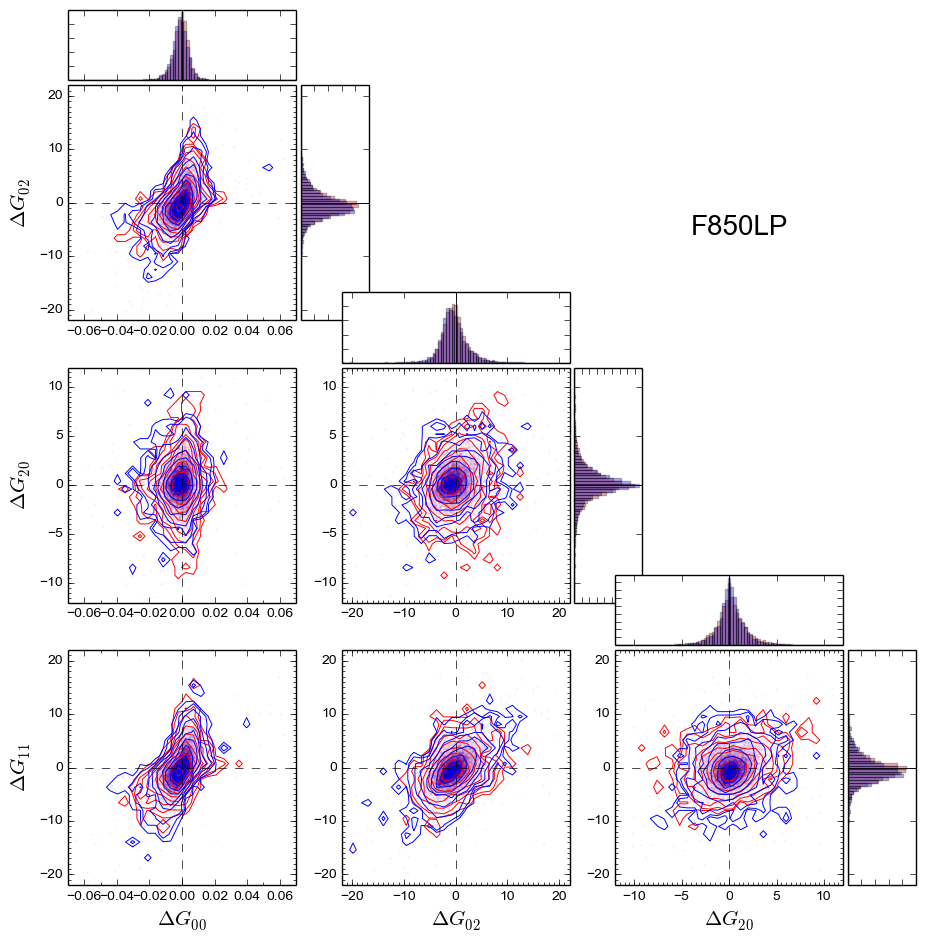}
  \caption {Dispersion around the true value of the brightness moments of
    the models reconstructed with EMPCA (blue) and
    as measured with Shapelens (red).}
  \label {fig:moments}
\end{figure*}

Figure~(\ref{fig:models}) displays four galaxies belonging to the same
four samples used to create the basis shown in
Fig.~(\ref{fig:basis}). The images have been rescaled to better
visualize their details. For each object we show, from left to right,
the original image, the model, the residuals, and the segmentation
used to remove nearby objects. The color images of the same galaxies
are shown in the top left and central panels of
Fig.~(\ref{fig:color-stamps}) together with additional examples. The
color stamps have a field size of $5.4$ arcsec, are based on the
F435W, F606W, and F775W filters and show once more the range of sizes
and morphologies which can be modeled. In this case, the galaxies are
visualized without any rescaling. The residuals are compatible with
the image noise except for very concentrated features which are
slightly missed because of the regularization scheme we adopted when
constructing the basis functions which the models are based on (see
Fig.~\ref{fig:models}).

\section {Testing the brightness moments}\label{sec:simulations}

In this section, we ``feed'' {\it EasySky} with the postage-stamps of
the galaxy models created in Sec.~(\ref{sec:models}) and extracted
from the XDF Survey with {\it EasyEMPCA}. In this case, the galaxies
have been arranged on a regular grid, randomly flipped, and rotated by
multiples of $90$ degrees. Their flux has not been changed to
produce an image as close as possible to the original data. Such image
permutations do not affect the original quality of the stamps because
no interpolation is involved in this process, as it would happen
when applying arbitrary rotations.

In Fig.~(\ref{fig:comparison-clean-dirty}), we show a realization of a
portion of $0.65 \times 0.65$ arcmin$^2$ field-of-view in the F755W
band. The noise-free image is shown in the top left panel, while the
other panels show three images with different levels of noise and
resolution to resemble the XDF survey (upper right panel), one orbit
exposure with HST (bottom left), and a CFHTLens stacked image with a
seeing of $0.7$ arcsec (bottom right panel). We can now process these
simulations like real images to verify and quantify the accuracy of
the galaxy reconstructions. Do do so, we first detect the objects with
SExtractor and then apply EasyPCA with the complete procedure to
derive the basis and the models (see Sec.~\ref{sec:empca}). It is
important to note that even if the mock galaxies used in the
simulations have been produced with the EMPCA, we took care to create
a sample of galaxies which is as independent as possible from the
original sample. This is why the galaxy images have been randomly
flipped, rotated and split into training sets differing, in number and
components, from the training set used to create the simulations in
first place. Further more, the noise in the simulated images is not
the same as that in the real data.

To have a quantitative assessment of the reconstruction quality, we
measured the brightness moments,
\begin{equation}
  G_{ij} = \int_{-\infty}^\infty d(\vec{x}) x_1^i x_2^j \d x \;,
\end{equation}
of all galaxies in the simulation with noise to quantify the
deviations with respect to those expected in the noise-free images. In
Fig.~(\ref{fig:moments}), we plot the scatter in the brightness
moments, $\Delta G_{ij}=G_{ij}^{measured}-G_{ij}^{true}$, up to the
second order (as required for weak-lensing measurements). The moments
except $G_{00}$ have been normalized by flux.

We finally processed the same images with a well-established
weak-lensing method to have a direct comparison. At this end, we used the
Shapelens library which allows to measure the brightness moments by
iteratively matching the data to an elliptical weight function to
maximize the signal-to-noise ratio \citep{2011MNRAS.412.1552M}. The
brightness moments of the models based on the EMPCA (blue contours)
are reproduced with an accuracy comparable to that achieved by
Shapelens (red contours), proving the quality of the models.

\section{Conclusions}\label{sec:conclusions}

We have described how optical images of galaxies can be fitted with an
optimized linear model based on the Expectation Maximization Principal
Components Analysis (EMPCA). This method relies on the data alone,
avoiding any assumptions regarding the morphology of the objects to be
modeled even if they have complex or irregular shapes. As a test case,
we have analyzed the galaxies listed in the \cite{rafelski15} catalog
which covers the Hubble eXtreme Deep Field (XDF). We selected those
objects with magnitude $m_{W755}<30$, far from the field edges and
without overlapping artifacts caused by the few stars present across
the field. We collected 7038 postage-stamps of noise-free galaxy
images with redshift up to $z=4.0$.

We have shown how the modeled galaxies well represent the entire
collection of galaxies, from small to large and from regular to
irregular. Two codes have been implemented to this end: {\it EasySky}
to create the simulations and {\it EasyEMPCA} to model the
galaxies. The residuals appear uncorrelated except at very sharp
features because of the regularization scheme we adopted during the
basis construction. To further verify the quality of the
reconstructions, we simulated a set of galaxy images, with and without
noise, covering the entire spectrum of shapes and luminosities of the
objects present in the XDF. We processed the simulations with the same
procedure applied to a real data set: we detected the objects with
SExtractor, derived the EMPCA basis and fitted the data with the
linear model based on this basis. We then measured the brightness
moments up to the second order of the model reconstructions and
compared them to those of the noise-free simulations. The quality of
the reconstructions very well competes with a well-established method
to measure galaxy brightness moments such as the iterative adaptive
scheme implemented in Shapelens.

The procedure discussed in this paper can be used to derive the
properties of galaxies such as their fluxes and shapes, or to create
reliable simulations of optical images. In this respect, the accuracy
of such simulations is gaining importance for the lensing
community. For instance, in the strong-lensing regime, they are
necessary to understand how substructures in strongly magnified
galaxies can be used to access additional information on the lensing
mass distribution, such as galaxy clusters. In the weak-lensing
regime, all methods to measure the ellipticites of galaxies require
precise simulations for their calibration, on which depends the bias
of such measurements and all quantities derived from them. The method
we discussed in this work appears as a promising solution to create
such simulations.

\section*{Acknowledgments}

We would like to thank Marc A. Rafelski and Anton M. Koekemoer for
kindly provide us the segmentation map of the UVUDF data, and Massimo
Meneghetti for the useful discussion. This work was supported by the
Transregional Collaborative Research Centre TRR 33.


\end{document}